\documentclass[conference]{IEEEtran}
\usepackage{amsmath}
\usepackage{amssymb}                                           
\usepackage{MnSymbol}
\usepackage{amsxtra,amsthm}
\usepackage{bbold}
\usepackage{mathdots}
\usepackage{mathtools}
\usepackage{pdfcolmk}
\usepackage{graphicx}                                        
\usepackage{algorithm}
\usepackage{algorithmic} 
\usepackage{mathrsfs}
\usepackage{todonotes}

\usepackage[utf8]{inputenc}                                     
\usepackage[T1]{fontenc}
\usepackage{verbatim}
\usepackage{csquotes}
\usepackage{framed} 
\usepackage{tikz,pgfplots}
\usepackage{pgfpages}
\usepackage{filecontents}
\usetikzlibrary{positioning,decorations.pathreplacing}

\usepackage{srcltx}
\usepackage[english]{babel}

\usepackage[shortlabels]{enumitem}
\usepackage[all]{xy}

\usepackage[style=ieee,    
            firstinits=true, 
            url=false,
            backend=bibtex,
            doi=false,
            maxbibnames=99,
            isbn=false,
            natbib=true,
            hyperref=false,
            bibencoding=utf8]{biblatex}
\AtEveryBibitem{\clearfield{note}}   
\DefineBibliographyStrings{english}{%
  references = {},
}

\bibliography{abbrv,jabref_philipp_utf2,peter}

\newcommand{\mgeq}{\succeq}

\newcommand{\Lone}{{{L}_1}}
\newcommand{\Ltwo}{{{L}_2}}

\newcommand{\norm}[1]{\left\|#1\right\|}

\let\forallalt\forall
\renewcommand{\forall}{\;\forallalt\;}

\let\refalt\ref
\renewcommand{\ref}[1]{(\refalt{#1})}

\renewcommand{\vec}{\mathbf}

\newcommand{\del}{\ensuremath{\delta}}

\newcommand{\Gam}{\ensuremath{\Gamma}}

\newcommand{\tht}{\ensuremath{\theta}}

\newcommand{\vGam}{\ensuremath{\mathbf{\Gam}}}

\newcommand{\vdel}{\ensuremath{\boldsymbol \delta}}

\newcommand{\va}{{\ensuremath{\mathbf a}}}
\newcommand{\vb}{{\ensuremath{\mathbf b}}}

\newcommand{\vh}{{\ensuremath{\mathbf h}}}                         
\newcommand{\vn}{{\ensuremath{\mathbf n}}}                         

\newcommand{\vr}{{\ensuremath{\mathbf r}}}                         

\newcommand{\vx}{{\ensuremath{\mathbf x}}}

\newcommand{\vxone}{{\ensuremath{\mathbf x}_1}}
\newcommand{\vxtwo}{{\ensuremath{\mathbf x}_2}}

\newcommand{\vy}{{\ensuremath{\mathbf y}}}

\newcommand{\Rvx}{{\ensuremath{\mathbf x}^{-}}}
\newcommand{\Rvxj}{{\ensuremath{\mathbf x}^{-}_j}}

\newcommand{\Rvxone}{\ensuremath{{\mathbf x}_1^{-}}}
\newcommand{\Rvxtwo}{{\ensuremath{\mathbf x}_2^{-}}}

\newcommand{\vth}{\ensuremath{\tilde{\mathbf h}}}

\newcommand{\vtn}{\ensuremath{\tilde{\mathbf n}}}
\newcommand{\vtx}{\ensuremath{\tilde{\mathbf x}}}

\newcommand{\hvx}{{\ensuremath{\hat{\mathbf x}}}}

\newcommand{\uA}{{\ensuremath{\mathrm A}}}

\newcommand{\uX}{{\ensuremath{\mathrm X}}}

\newcommand{\uY}{{\ensuremath{\mathrm Y}}}

\newcommand{\uH}{{\ensuremath{\mathrm H}}}

\newcommand{\vA}{{\ensuremath{\mathbf A}}}

\newcommand{\vS}{\ensuremath{\mathbf S }}                         
\newcommand{\vT}{\ensuremath{\mathbf T}}

\newcommand{\vX}{{\ensuremath{\mathbf X}}}

\newcommand{\zero}{{\ensuremath{\mathbf 0}}}

\newcommand{\vzero}{{\ensuremath{\mathbb 0}}}

\newcommand{\tx}{\ensuremath{\tilde{x}}}

\newcommand{\tL}{\ensuremath{\tilde{L}}}

\newcommand{\tN}{\ensuremath{\tilde{N}}}

\newcommand{\htx}{\ensuremath{\hat{\tx}}}

\newcommand{\Calg}{\ensuremath{\mathscr C}}\newcommand{\Code}{\Calg}

\newcommand{\Alin}{{\ensuremath{\mathcal{A}}}}

\newcommand{\C}{{\ensuremath{\mathbb C}}}

\newcommand{\Fmatrix}{{\ensuremath{\mathbf F}}}

\newcommand{\Fmatrixa}{{\ensuremath{\mathbf F}^*}}
\newcommand{\FmatrixN}{{\ensuremath{\mathbf F}_N}}

\newcommand{\Zform}{{\ensuremath{\mathcal Z}}}

\newcommand{\id}{{\ensuremath{\mathbb 1}}}

\newcommand{\eins}{{\ensuremath{\mathbf 1}}}

\newcommand{\Pro}{\prod}
\newcommand{\Proj}{{\mathbf \Pi}}
\newcommand{\ProjNL}{\ensuremath{{\mathbf \Pi}_{N,L}}}

\newcommand{\ProjNLj}{\ensuremath{{\mathbf \Pi}_{N,L_j}}}

\newcommand{\skprod}[2]{\ensuremath{ \left\langle #1,#2 \right\rangle }}

\definecolor{gray}{rgb}{0.3,0.3,0.3}
{\color{black}}                      
{\color{black}}

{\color{black}}                      
{\color{black}}

\newcommand{\thmref}[1]{Theorem~\ref{#1}}     

\newcommand{\secref}[1]{Section~\ref{#1}}

\newcommand{\figref}[1]{Figure~\ref{#1}}

\newcommand{\noi}{\noindent}

\ifx\definition\undefined
\newtheorem{definition}{Definition}         
\fi
\ifx \@definition \@empty
\newtheorem{definition}[theorem]{Definition}   
\fi
\ifx\corrolary\undefined
\newtheorem{corrolary}{Corrolary} 
\fi
\ifx\conjecture\undefined
\newtheorem{conjecture}{Conjecture}         
\fi
\ifx\theorem\undefined
\newtheorem{theorem}{Theorem}         
\fi
\ifx\lemma\undefined
\newtheorem{lemma}{Lemma}
\fi
\ifx\question\undefined
\newtheorem{question}{\color{red}{Question}}         
\fi

\newenvironment{remark}{\par\vspace{1.5ex}\noindent{\em Remark\/}.}{\par\vspace{1.5ex}}

\DeclareMathOperator{\find}{find}

\newcommand\tr{\operatorname{tr}} 

\newcommand{\argmin}[1]{\underset{#1}{\operatorname{argmin}}}

\newcommand{\Norm}[1]{\ensuremath{ \left\|#1\right\| }}

\newcommand{\cc}[1]{{\ensuremath{\overline{#1}}}} 

\makeatletter
  \newcommand{\set}[2]{\ensuremath{%
  \setbox0=\hbox{\ensuremath{#2}}
  \dimen@\ht0
  \advance\dimen@ by \dp0
  \left\{\left.#1\rule[-\dp0]{0pt}{\dimen@}\;\right|\;#2\right\} }}
\makeatother

\renewcommand{\argmin}{\operatornamewithlimits{argmin}}

\newcommand{\namen}[1]{{\textsc{#1}}}           

\ifx \@paragraph \@empty
\makeatletter
\renewcommand\paragraph{\@startsection
{paragraph}{4}{\z@}{-3.5ex plus-1ex minus-.2ex}%
{1.3ex plus.2ex}{\normalfont\itshape}}
\makeatother
\fi

{\par\noindent{\em Beweis\/}.}%
{\hspace*{\fill}{\qed}\vspace{1ex}\par}
{\par\noindent{\em Proof\/}.}%
{\par}

{\hspace*{\fill}{\lightning}\vspace{1ex}\par}
{\par\vspace{1.5ex}\noindent{\em Remark\/}.}
{\par\vspace{1.5ex}}
\ifx\remark\undefined
\newenvironment{remark}%
{\par\vspace{1.5ex}\noindent{\em Remark\/}.}
{\par\vspace{1.5ex}}
\fi
\ifx\hremark\undefined
\newenvironment{hremark}%
{\par\vspace{1.5ex}\noindent{\em Historic Remarks\/}.}
{\par\vspace{1.5ex}}
\fi
{\par\vspace{1.5ex}\noindent{\em Recall\/}.}
{\par\vspace{1.5ex}}
\newenvironment{beispiel}%
{\par\vspace{1.5ex}\noindent{\em Example\/}. }
{\par\vspace{1.5ex}}
\ifx\openproblem\undefined
\newenvironment{openproblem}%
{\par\vspace{1.5ex}\color{brown}\noindent{\em Open Problem\/}. }
{\par\vspace{1.5ex}\color{black}}
\fi
\ifx\open\undefined
\newenvironment{open}%
{\par\vspace{1.5ex}\color{brown}\noindent{}. }
{\par\vspace{1.5ex}\color{black}}
\fi

{\noi\vspace{0.5ex}\small}
{\vspace{0.5ex}\par\normalsize}

\newif\ifdetails\detailstrue 
\detailsfalse 
\ifdetails
\newcommand{\detail}[1]{\color{brown}{#1}\color{black}}
\else
\newcommand{\detail}[1]{}
\fi

\newcounter{Examplecount}
{\renewcommand{\labelenumi}{(\roman{enumi})}\begin{list}{\labelenumi}
{\usecounter{enumi}\setlength{\labelwidth}{1.5cm}\setlength{\topsep}{0.3cm}\setlength{\itemsep}{-3pt}}}
{\end{list}}
{\renewcommand{\labelenumi}{(\arabic{enumi})}\begin{list}{\labelenumi}
{\usecounter{enumi}\setlength{\labelwidth}{1.5cm}\setlength{\topsep}{0.3cm}\setlength{\itemsep}{-3pt}}}
{\end{list}}
{\renewcommand{\labelenumi}{$\bullet$}\begin{list}{\labelenumi}
{\setlength{\labelwidth}{1.5cm}\setlength{\topsep}{0.3cm}\setlength{\itemsep}{-2pt}}}
{\end{list}}

\makeatletter 
  \newcommand{\sprod}[2]{\ensuremath{%
  \setbox0=\hbox{\ensuremath{#2}}
  \dimen@\ht0
  \advance\dimen@ by \dp0
  \left(\left.#1\rule[-\dp0]{0pt}{\dimen@}\right|#2\right)}}
\makeatother 

\newcommand{\seefor}[1]{\!}    
\newcommand{\seeintern}[1]{\!} 

\begin{document}
%
%
\title{Short-Message Communication and FIR System Identification using Huffman Sequences}

\author{
\IEEEauthorblockN{Philipp Walk\IEEEauthorrefmark{1}, Peter Jung\IEEEauthorrefmark{2}, and Babak Hassibi\IEEEauthorrefmark{1}}
\IEEEauthorblockA{\IEEEauthorrefmark{1}Department of Electrical Engineering, Caltech, Pasadena, CA 91125\\
Email: \{pwalk,hassibi\}@caltech.edu}
\IEEEauthorblockA{\IEEEauthorrefmark{2}Communications \& Information Theory, Technical University Berlin, 10587 Berlin\\
Email: peter.jung@tu-berlin.de}
}

\maketitle
\begin{abstract} 
  Providing short-message communication and simultaneous channel estimation for sporadic and fast fading scenarios is a
  challenge for future wireless networks.  In this work we propose a novel blind communication and deconvolution scheme by using Huffman
  sequences, which allows to solve three important tasks in one step: (i) determination of the transmit power (ii)
  identification of the discrete-time FIR channel by providing a maximum delay of less than $L/2$ and (iii)
  simultaneously communicating $L-1$ bits of information.  Our signal reconstruction uses a recent semi-definite program
  that can recover two unknown signals from their auto-correlations and cross-correlations.  This convex algorithm is
  stable and operates fully deterministic without any further channel assumptions.
\end{abstract}

\section{Introduction}

Next generation wireless communication networks have to cope simultaneously with many different and partially
contradicting tasks. It becomes increasingly apparent that current technologies will not be able to meet the emerging
demands of future mobile communication systems, such as supporting sporadic and short-message traffic types for the
internet of things, machine--type communication and sensor applications.  In particular, once a node wakes up in a
sporadic manner to deliver a message it has first to indicate its presence to the network. Secondly, training symbols
(pilots) are used to provide sufficient information at the receiver for estimating link parameters such as the channel.
Finally, after exchanging a certain amount of control information the device transmits its desired information message
on pre-assigned resources. In current systems these steps are usually performed in separate communication phases
yielding a tremendous overhead once the information message is sufficiently short and the nodes wake up in an
unpredictable way.  Therefore, a redesign and rethinking of several well-established system concepts and dimensioning of
communication layers is necessary to support such traffic types in an efficient manner \cite{Wunder2015:sparse5G}.  This
has gained again deeper interest in methods for blind demixing and deconvolution methods which can operate on a
short-frame basis.  Disadvantages of the classical techniques hereby lie in its statistical flavour and in the lack of
efficiency and robustness, since the algorithms for identification are often iterative and rarely have convergence
guarantees.  

In this work we will use a convex program for the channel and data reconstruction first introduced in \cite{JH16} for
the noise free case and show its numerical stability. The blind reconstruction can hereby be re-casted as a phase
retrieval problem with additional knowledge of the auto-correlations of the data and the channel at the receiver.  The
uniqueness of the phase retrieval problem can then be shown by constructing an explicit dual certificate in the noise
free case, as was shown in \cite{JH16} for almost all signals and channels.  In \cite{WJPH16a} and more detailed in
\cite{WJPH17} we have shown, that the uniqueness derived in \cite{JH16}, holds indeed deterministically, given a
particular co-prime condition is fulfilled. The latter condition was already shown in \cite{XLTK95} to be necessary for
blind deconvolution. Using therefore sequences with good autocorrelation properties allows to estimate the
autocorrelation of the channel from the observations which in turn enables blind deconvolution by solving the
corresponding phase-retrieval problem.  Here, we propose to use Huffman sequences for this purpose which comes with
further advantages from the system identification perspective.  This scheme allows for simultaneous FIR channel
estimation, transmit power estimation, to resolve near-far effects, and the communication of short messages.

The paper is organized as follows: First we motivate and introduce in \secref{sec:blinddeconv} deterministic blind
deconvolution with additional knowledge of autocorrelations for a short-message communication. Then, in
\secref{sec:goodautocorrelation}, we investigate signal classes of good and known autocorrelations, yielding to a
codebook design of Huffman sequences. Due to the impulsive-equivalent behaviour of their autocorrelations, we show in
\secref{sec:blindhuff}, that they can be used to obtain a good estimate  of the channel autocorrelation at the receiver.
By using the SDP in \cite{JH16} we show a perfect reconstruction of channel and data in the noise free case. Finally, in
\secref{sec:simulation}, we demonstrate numerically noise robustness of our reconstruction scheme in terms of
bit-error-rates.  

\section{Blind Deconvolution}\label{sec:blinddeconv}

One-dimensional blind deconvolution problems occur in many signal processing applications, such as in digital communication over wired
or wireless channels, where the channel is modeled as a linear time invariant (LTI) system, which has to be blindly
identified or estimated. 
\begin{equation}
  y_k=(\vec{h}\ast \vec{x})_k=\sum_{l}h_lx_{k-l}\label{eq:convtime}.
\end{equation}
If the receiver has some statistical knowledge of the LTI system, as given for example by second or higher order
statistics, known blind channel equalization and estimations were already developed in the $90$'s (see for example in
\cite{TXK91,DKAJ91,TXHK95}).  If no statistical knowledge of the data $\vx$  and the channel $\vh$ is available, for example, for
fast fading channels, one can still ask under which conditions on the data and the channel a blind channel or system
identification is possible. Necessary and sufficient conditions in a multi-channel setup where first derived in
\cite{XLTK95} and continuously further developed (see e.g.  \cite{A-MQH97} for a nice summary).  However, the desired
application here is a single--channel blind deconvolution of short signals (frame length is in the order of $1,2,\dots$
times the maximum delay spread of the channel), whereas previous methods often fail.

Recent progress in low--rank matrix recovery have put the one-shot blind deconvolution as a prototypical bilinear
inverse problem back into focus. Using lifting, new results are obtained in \cite{Ahmed:2012} for randomized cyclic
convolutions. There, the data signal lies in a random low--dimensional subspace and with certain incoherence assumptions
it can be recovered with high probability using nuclear norm minimization.  The computational aspects of the unlifted
problem has been tackled recently in \cite{Li2016} with a clever initialization overcoming with high probability the
non-convex nature such that gradient based algorithms will not stuck in a local minima.  Although this renders blind
deconvolution tractable in theoretical terms this (i) requires cyclic extensions which can be itself in the order of the
signal length for our desired application, (ii) it requires common knowledge on random parameters which is often not
feasible and (iii) even in the noiseless setting the recovery guarantees are probabilistic and it is therefore difficult
to fulfill strict system requirements.

We will therefore address in this context blind (aperiodic) deconvolution again in the classical framework of
polynomial factorization. Here, the convolution \eqref{eq:convtime} transfers with the $z-$transform $\Zform$, given for any
$\vx\in\C^L$ as
\begin{align*}
  \Zform\ \colon \C^L \to \C[z], \quad 
  \vx\mapsto(\Zform\vx)(z)=\uX(z):=\sum_{k=0}^{L-1} x_k z^{-k}
\end{align*}
to a polynomial multiplication 
\begin{equation}
  \vspace*{-.0em}
  \uY(z)=\uH(z)\cdot \uX(z)=\sum_{k=0}^{L+K-2} y_k z^{-k}.\label{eq:uY}
  \vspace*{-.0em}
\end{equation}
Note, $\uY, \uH,$ and $\uX$ are polynomials in the variable $z^{-1}\in\mathbb{C}$.  Hence, given the
observation $\vec{y}$ and further constraints/knowledge of $\vec{h}$ and $\vec{x}$, the recovery problem is equivalent
to find the factorization \eqref{eq:uY}. Indeed, the program in \cite{JH16} obtains the right unique factorization by
additionally knowledge of the autocorrelations $\uH\uH^*$ and $\uX\uX^*$ of the factors, which have to be co-prime. 

\section{Good Autocorrelation Sequences}\label{sec:goodautocorrelation}

Blind deconvolution using the SDP proposed in \cite{JH16} and \cite{WJPH16a,WJPH17} is based on the idea that one has
access to the autocorrelations of the transmitted data signal and of the channel. For any $\vx\in\C^L$ the autocorrelation
$\va_x:=\vx*\cc{\vx^-}$ is defined by the convolution of $\vx$ with its conjugate-time-reversal given by
$(\cc{\vx^-})_k=\cc{x_{L-1-k}}$ for $k\in\{0,\dots,L-1\}$. We will explain this program  later in
Section~\ref{sec:blindhuff}.  In the desired application, however, the receiver has only access to the observed channel
output. But, if we a-priori fix the autocorrelation $\va_x$ of the data, the autocorrelation of the channel $\va_h$ can
be estimated from the channel output and we can use the ambiguities of $\va_x$ for communicating a short message in $\vx$. Let
us illustrate this in the noiseless setting.  Using the \namen{Wiener-Lee} relations:
\begin{align}
  \va_y=\vy*\cc{\vy^-}=(\vx*\cc{\vx^-})* (\vh*\cc{\vh^-})=\va_x * \va_h\label{eq:wienerlee},
\end{align}
see e.g. \cite[(2.29)]{Lue92}, we can retrieve $\va_x$ from $\va_y$ if $|\uA_x(z)|>0$ on the unit circle. It is
obvious that this is possible and sufficiently stable if the convolution behaves close to an identity for the desired
channels (we will assume that the maximum delay spread is known). In other words, to obtain from the received signal
$\vy=\vx*\vh$ the autocorrelation of $\vh$, we need further properties of $\vx$, in the sense that the autocorrelation
is close to an impulse.

\subsection{Huffman Sequences}
\begin{figure}[t]
  \begin{minipage}{0.495\columnwidth}
    \includegraphics[width=\columnwidth]{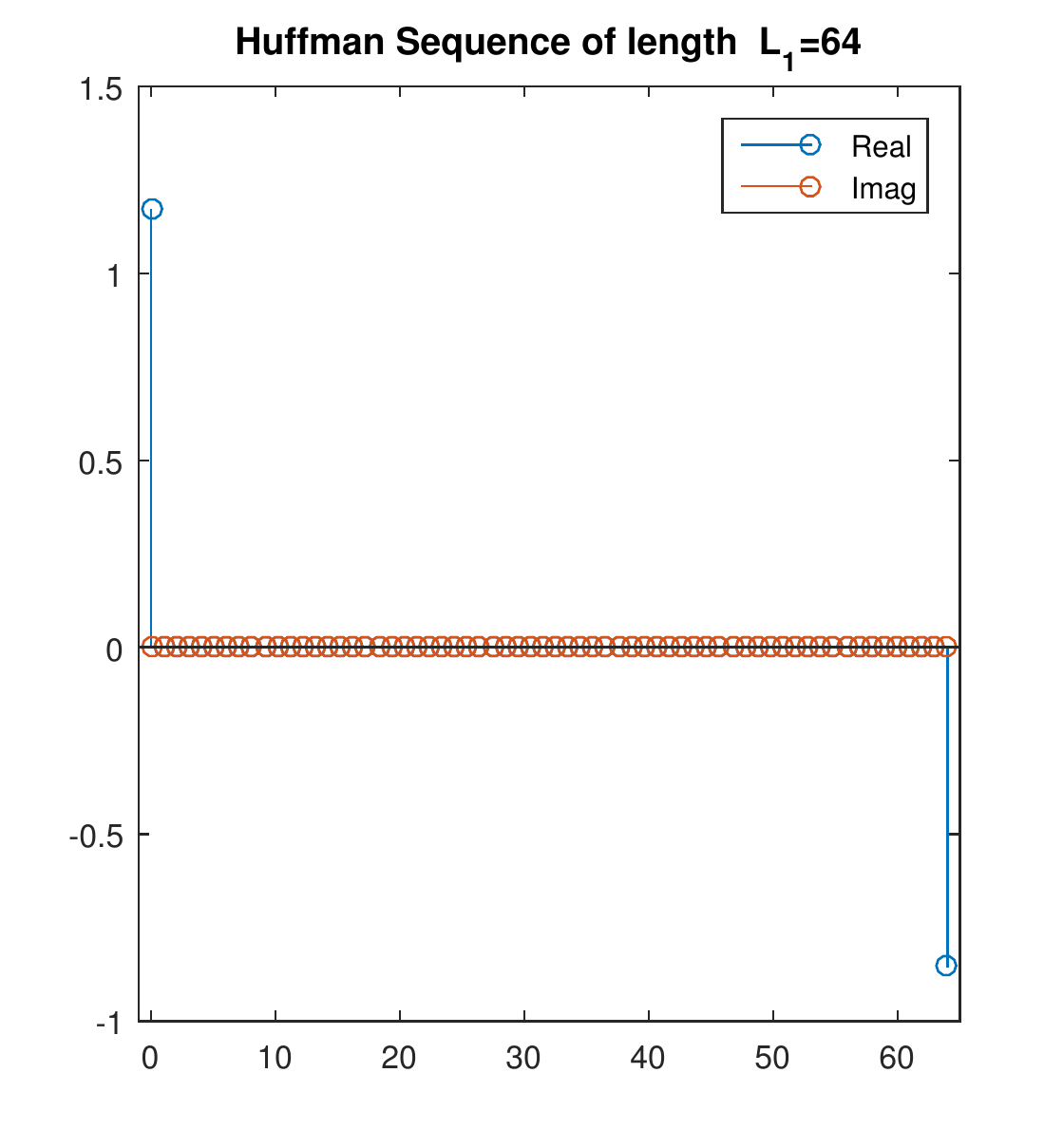}
    \caption{Huffman codeword in time for all one bit, $L=64$ and $E=2.1$.}\label{fig:x1x2ambi}
  \end{minipage}
  \begin{minipage}{0.495\columnwidth}
    \includegraphics[width=\columnwidth]{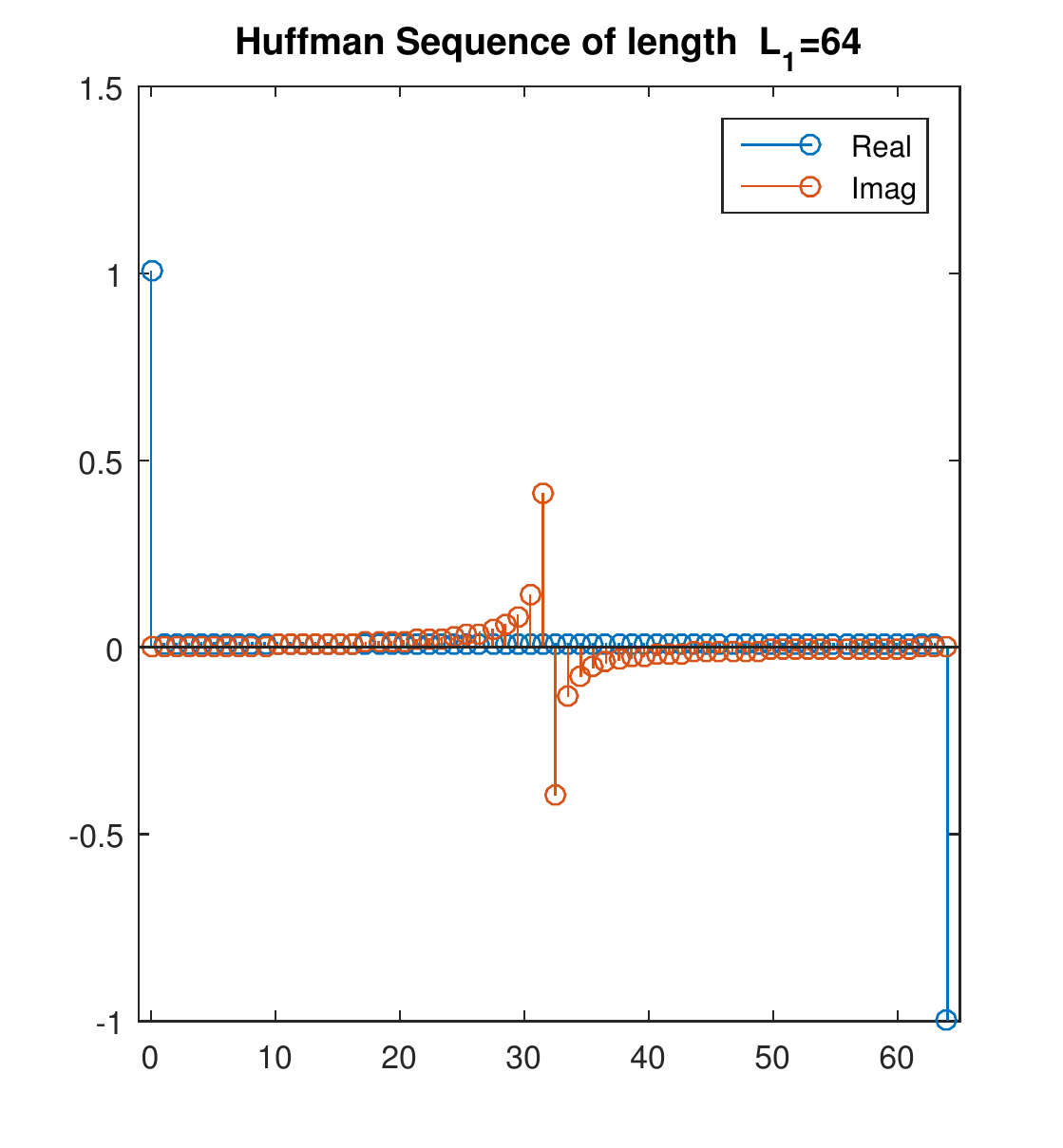}
    \caption{Huffman codeword in time for same amount of ones and zeros.}\label{fig:sameonezero}
  \end{minipage}
\end{figure}

For the cyclic (periodic) autocorrelation of sequences (vectors) $\vx\in\C^{L}$, having the impulse-vector\footnote{
Given by the Kronecker tensor $(\vdel_0)_k=\del_{0,k}=0$ for $k\not=0$ and $\del_{0,0}=1$.}
$\vdel_0$ as autocorrelation, are  called \emph{perfect sequences}, see e.g. \cite[5.8]{Lue92}.
Unfortunately, for aperiodic autocorrelations $\va=\va_x=\vx*\cc{\vx^-}$, it is easily seen that a perfect aperiodic
autocorrelation can not exist if $\vx\not=\vdel_0$, since for any
$\vx\in\C_{0,0}^L:=\set{\vx\in\C^L}{x_0\not=0\not=x_{L-1}}$ with\footnote{
If $L=1$ we get the trivial multiplication $a=x\cc{x}=|x|^2=1$, having only a global phase solution $e^{i\phi}$.}
$L\geq 2$ we obtain for the first and last coefficient
\begin{align}
  a_{0}=x_0 \cc{x_{L-1}} = \cc{x_{L-1}\cc{x_0}}=\cc{a_{2L-2}}\not=0.\label{eq:sidelobmin}
\end{align}
Nevertheless, there exists a huge literature on constructing almost perfect aperiodic autocorrelation sequences, see for
example \cite[Cha.6]{Lue92}.  Since our goal is to use them for identifying the channel autocorrelation we will need
impulsive ones, i.e., where most of the sidelobes vanish. In fact, the best of such impulse like autocorrelations were
found by \namen{Huffman} \cite{Huf62} and are given by
\begin{align}
  a_k = \begin{cases}
    -1 & k=0\\ 
    0 & k\in \{1,2, \dots, L-2\}\\
    E & k=L-1\\
    0 & k\in \{L,L+1, \dots, 2L-3\}\\
    -1 & k=2L-2 
  \end{cases}\label{eq:huffmanautocor}
\end{align}
with energy $E=\norm{\vx}^2\geq |x_0|^2+|x_{L-1}|^2 \geq 2$. 
The construction of such sequences is straight-forward in the $z-$domain by determining its zero.
Following the lines in \cite{BA91} we get for the autocorrelation in the $z-$domain
\begin{align}
  \uA(z)=\uA_x(z) = -1 + E z^{-(L-1)} -z^{-2(L-1)},\label{eq:huffmanz}
\end{align}
which is a polynomial of order $2L-2$, having $2L-2$ zeros $\zeta\in\C$. Indeed, if $\zeta^{-L+1}=r>0$ then
$\zeta$ is a zero of $\uA(z)$ for 
\begin{align}
  r_{\pm}= \frac{E}{2} \pm \sqrt{\frac{E^2}{4}-1},
\end{align}
which implies $r_+=r_-^{-1}\geq 1$.  Hence, the $2L-2$ zeros have radius $R_{\pm}:=r_{\pm}^{1/(L-1)}$  and are given for
$k\in \{1,\dots, L-1\}$ by
\begin{align}
  \zeta_k^{\pm} := e^{-2\pi i\frac{k-1}{L-1}}\cdot R_{\pm} .\label{eq:huffmanzeros}
\end{align}
\detail{To see that $r_+=r_-^{-1}$, we just have to show $r_+ r_-=1$, which holds by 
  \begin{align}
    r_+ r_- &= \left(\frac{E}{2} + \sqrt{\frac{E^2}{4} -1}\right)\left(\frac{E}{2} - \sqrt{\frac{E^2}{4}
  -1}\right)\\
  &= \frac{E^2}{4} -\left( \frac{E^2}{4} -1\right)=1.
\end{align}
}
Since $\zeta_k^+=1/\cc{\zeta_k^-}$, the $2L-2$ zeros occur in conjugated-pairs, where $L-1$ zeros lie on the circle of radius
$R_+$ and the other $L-1$ on the circle of radius $R_{-}$, i.e.,
\begin{align}
  \uA(z)=-\Pro_{k=1}^{L-1} (z^{-1} -\cc{\zeta_k^{-}})\cdot\Pro_{k=1}^{L-1} (z^{-1}-\cc{\zeta_k^+})=\uX(z)\uX(z)^*. \label{eq:Az}
\end{align}
Note, we have to set the unit (scaling) to $-1$, since that the last coefficient becomes $-1$ and the first 
$-\Pro_{k=1}^{L-1} e^{4\pi k/(L-1)}=-1$, if we calculate the product in  \eqref{eq:Az}.  By swapping the primes, i.e., the
zeros, we can obtain $2^{L-1}$ different factorizations, the maximal amount of non-trivial ambiguities of the
autocorrelation \cite{WJPH17}, yielding by the inverse $z-$transform to $2^{L-1}$ different Huffman sequences
$\Zform^{-1}\uX=\vx\in\C^L_{0,0}$ having all the same energy $E$ and autocorrelation \eqref{eq:huffmanautocor}. Since
$\uX$ is up to a unit $e^{i\tht}$ defined by its zeros, we can set the unit to $1$ which yields as first coefficient
$x_0=1$ and hence as last coefficient $x_{L_1-1}=-1$, see \figref{fig:sameonezero}. But if we assign
$B_+\subseteq\{1,\dots,L-1\}$  with $|B_+|\not=(L-1)/2$ (always true if $L$ is even) zeros $\zeta_k^+$ for $\uX$, then
we have to assign $|B_-|=L-1-|B_+|$ zeros of radius $R_-$ with $B_-=B_+^c$, which gives $\Pro_{k\in B_+}
\cc{{\zeta}_{k}^{-}}\Pro_{B_-}\cc{\zeta_k^+}=R_+^{|B_-|-|B_+|}=:c^2\not=1$. Hence, we have to scale our selection by
\begin{align}
  \uX(z):= -c \Pro_{k\in B_+} (z^{-1} - \cc{\zeta_{k}^{-}})\Pro_{k\in B_-} (z^{-1} -
  \cc{\zeta_{k}^{+}}),\label{eq:uX}
\end{align}
where we assume $L$ even, yielding always a positive scaling factor\footnote{The product is real-valued since
$\Pro_{k=1}^{L-1} e^{-2\pi i (k-1)/(L-1)}=e^{2\pi i (\sum_k k)/(L-1)}=e^{\pi i L} =1$ if $L$ is even and $-1$ if $L$ is
odd.}
and therefore to $x_0>0$ and $x_{L-1}<0$. 
%
Then indeed, forming the involution of $\uX$  gives
\begin{align}
  \uX^*(z)&:= -z^{-L} \cc{\uX(1/\cc{z})}
  = -c\Pro_{k\in B_+} (1-\zeta_k^{-} z^{-1}) \Pro_{k\in B_-} (1-\zeta_k^{+} z^{-1})\notag\\
  &= c \Pro_{k\in B_+} \cc{\zeta_k^+}\Pro_{k\in B_-}\cc{\zeta_k^-} \Pro_{k\in B_+} (z^{-1}-\zeta_k^{-}) \Pro_{k\in
  B_-} (z^{-1}-\zeta_k^{+}) \notag 
\end{align}
and hence $\uX(z)\uX^*(z) = \uA(z)$.
\paragraph*{Encoding rule} We have designed above a non-binary block code $\Code_{E,L}$ of complex Huffman sequences
$\vx=\Zform^{-1}\uX$ of length $L$ and cardinality $2^{L-1}$. This allows encoding of $L-1$ bits by its zeros
\eqref{eq:huffmanzeros}. For the $k$th bit $b_k$ we set then
\begin{align}
  \zeta_k^{\pm}=\begin{cases}  e^{-2\pi i\frac{k-1}{L-1}}R_+ &, b_k =1\\
    e^{-2\pi i\frac{k-1}{L-1}} R_- &, b_k = 0\end{cases}.\label{eq:encdoingrule}
\end{align}
Such a rule needs to be implemented efficiently. For example, by adjusting the phase in $c$ and the
main-sidelobe ratio, integer sequences can generated recursively for certain lengths $L$ \cite{Lue92}.   

{\em Comments on the Peak-to-Average Power Ratio (PAPR):}
One drawback in using Huffman sequences without further restriction, i.e., with the
autocorrelation given in \eqref{eq:huffmanautocor}, is that this comes 
with an PAPR 
\begin{align}
  \text{PAPR}_{\Code_{E,L}}: = \max_{\vx\in\Code_{E,L}} L_1\cdot\frac{\Norm{\vx}_{\infty}^2}{\Norm{\vx}_2^2} \leq L_1,
\end{align}
where for $E\gg 2$ the maximum is achieved for the all zero or all one bit codewords, having they energy located at the first and last
coefficient, see \figref{fig:x1x2ambi}. If we set $E=2$ we obtain the best possible PAPR of $L/2$, but loosing our code
structure since $R^+=R^{-}=1$. Exemplary, by choosing $E=2.1$ we obtain for $L=128$ an PAPR of $19$dB, which is slightly higher
than for OFDM. Reducing the signal length to $64$ only yields to
$16$dB, see Figure \ref{fig:x1x2ambi} and \ref{fig:sameonezero}.
To further reduce the PAPR extreme signals from the code have to excluded
, i.e., the one where the zeros are concentrated on one of the two
circles. 
%

\section{Blind Deconvolution and decoding of Huffman Sequences}\label{sec:blindhuff}

Let us assume we have a finite impulse response (FIR) channel $\vh\in\C^{K}_{0,0}$ of length $K\geq 1$ with non-vanishing first and last
coefficients. If we use Huffman sequences $\vx\in\C^{L}_{0,0}$ with length $L= 2K+M$ for some $M\geq 0$,
we  receive 
\begin{align}
  \vy:=\vx*\vh\in\C^{L+K-1}.
\end{align}
To apply \thmref{thm:4correlation} we need knowledge of the autocorrelation of $\vx$ and of $\vh$. Indeed, using the
\namen{Wiener-Lee} relations \eqref{eq:wienerlee}, we get from the autocorrelation of the received signal 
\begin{align}
  \va_y= \vy*\cc{\vy^-}=\vx*\cc{\vx^-} * \vh*\cc{\vh^-} =\va_x*\va_h\in\C^{2L+2K-3}.
\end{align}
Using the property \eqref{eq:huffmanautocor} of the Huffman sequences   
\begin{align}
  \va_y= \va_x*\va_h=\begin{pmatrix}
    -1 \\ \zero_{L-2} \\ E \\ \zero_{L-2}\\ -1\end{pmatrix} * \va_h = \begin{pmatrix} -\va_h \\\zero_M \\ E
    \va_h \\\zero_M\\
    -\va_h \end{pmatrix}= \begin{pmatrix}\va_{y,1} \\\zero_M\\ \va_{y,3}\\ \zero_M \\
    \va_{y,5}\end{pmatrix},\label{eq:aynoisefree}
\end{align}
we can determine the channel autocorrelation  by $\va_h=-\va_{y,1}$. Moreover, we can obtain from the ratio
\begin{align}
  E= {\Norm{\va_{y,3}}_2}/{\Norm{\va_{y,1}}_2}\label{eq:Eestimate}
\end{align}
the energy $E$ of the Huffman sequences. Hence, we have determined $\va_x$ and $\va_y$ exactly. 
Inserting both autocorrelations in \eqref{eq:fcp3N} yields up to a global phase $\phi$ the reconstruction 
\begin{align}
  \vx^{\#}=e^{i\phi}\begin{pmatrix} \vx \\ \cc{\vh^-}\end{pmatrix}\label{eq:reconstructionx}
\end{align}
and therefore the Huffman sequence $e^{i\phi}\vx$. Since for even $L$, the first coefficient $x_0>0$, we just have to
divide $\vx^{\#}$ by the phase of $x_0^{\#}$ and obtain the original $\vx$ and $\vh$. For $L$ odd $x_0<0$. 
\paragraph*{Decoding Rule} From the estimated codeword $\vx^{\#}$ we have to reverse \eqref{eq:encdoingrule} to obtain
the zeros $\zeta_n^{\#}$, from which we calculate its absolute value $R_n^{\#}$ and phase $\phi_n^{\#}$. Then, for
$k:=(L-1)\phi_n^{\#}/2\pi \mod L-1$ we set for the $k$th bit 
\begin{align}
  b_k =\begin{cases} 1 &, R_n^{\#}\geq 1\\
    0 &, R_n^{\#}<1\end{cases}
\end{align}
However, operationally this needs some further investigations for efficient sequence detection.  
\subsection{Deconvolution via Semi-Definite Programming} It is known, that the (aperiodic) \emph{autocorrelation}
of a vector signal $\vx\in\C^N$ does not contain enough information to obtain a unique recovery, see for example
\cite{WJPH17} or \cite{Hur89}, the idea is to use cross-correlation informations of the signal by partitioning $\vx$ in
two disjoint signals $\vx_1\in\C^{L_1}$ and $\vx_2\in\C^{L_2}$ with $N=L_1+L_2$, yielding $\vx^T=(\vx_1^T,\vx_2^T)$ if
stacked together.  To obtain $\vx$ is equivalent to solve a phase retrieval problem via a semi-definite program (SDP).
Here, the autocorrelation or equivalent the Fourier magnitude-measurements are represented as linear mappings on
positive-semidefinite rank$-1$ matrices, see \cite{JH16},\cite{Jag16}.  This is known as \emph{lifting}. The above
partitioning of $\vx$ yields then a block structure for the positive-semidefinite matrix $\vx\vx^*$.  
%
%
The linear measurement $\Alin$ is given component-wise by the inner products with the sensing matrices
$\vA_{i,j,k}$, defined below, which correspond to the $k$th correlation component of $\vx_i$ and $\vx_j$ for
$i,j\in\{1,2\}$.
Hence autocorrelation and cross-correlation can be obtain from the same object $\vx\vx^*$. 
Let us define the $N\times N$ down-shift and $N\times L$ embedding matrix as 
\begin{align}
  \vT_N=
  \begin{pmatrix} \zero_{N-1}^T  & 0 \\ \id_{N-1,N-1} & \zero_{N-1}
    \end{pmatrix}
    \quad\text{and}\quad
  \ProjNL& =\begin{pmatrix}
    \id_{L, L}\\
    \vzero_{N-L, L}\end{pmatrix}.
\end{align}
\seeintern{See also definition in \eqref{eq:projNL}}%
Then, the $L_i\times L_j$ rectangular shift matrices are defined as 
\begin{align}
  (\vT^{(k)}_{\!L_j,L_i})^T:=\Proj^T_{N,L_i}\vT^{k-L_j+1}_{N}\ProjNLj,
       \label{eq:LiLj}
\end{align}
for $k\in\{0,\dots,L_i+L_j-2\}=:[L_i+L_j-1]$, where we set $\vT^l_N:=(\vT^{-l}_N)^T$ if $l<0$. Then, the
\emph{correlation} between vectors of
dimensions $L_i$ and $L_j$ is given component-wise as 
\begin{align*}
  (\!\va_{i,j}\!)_k=(\vx_i*\cc{\Rvxj})_{k}&= \skprod{\vx_i}{(\vT^{(k)}_{\!L_j\!,L_i}\!)^T\vx_j}
  = \tr(\vT^{(k)}_{\!L_j\!,L_i}\vx_i\vx_j^*).
\end{align*}
\seeintern{See also \eqref{eq:discorr} and \eqref{eq:veccorr2}}%
Hence this defines the linear maps $\Alin_{i,j,k}(\vX):=\tr(\vA_{i,j,k}\vX)$, where $\vA_{i,j,k}$ are the 
sensing matrices given as the correspondingly  zero-padded $\vT_{L_j,L_i}^{(k)}$ in \eqref{eq:LiLj}.
Stacking all the $\Alin_{i,j}$ together gives finally the measurement map $\Alin$.
%
Hence, the $4N-4$ complex-valued linear measurements are
given by
\begin{align}
   \vb=\!\Alin(\vx\vx^*)\!=\!
        \Alin\begin{pmatrix}
        \vx_1\vx_1^* & \vx_1\vx_2^*\\
        \vx_2\vx_1^* & \vx_2\vx_2^*\end{pmatrix}\!
    =
    \begin{pmatrix} 
      \vxone*\cc{\Rvxone} \\ 
      \vxtwo*\cc{\Rvxtwo}\\
      \vxone*\cc{\Rvxtwo} \\
      \vxtwo*\cc{\Rvxone}
    \end{pmatrix}=\begin{pmatrix}\va_{1,1}\\\va_{2,2}\\\va_{1,2}\\\va_{2,1}\end{pmatrix}.\label{eq:3Nb}
\end{align}
Note, $\vb$ is {\bfseries not} an autocorrelation, but contains the part of the autocorrelation 
\begin{align}
  \begin{pmatrix} \vx_1\\ \zero_{L-1} \\ \vx_2\end{pmatrix}
  * \cc{\begin{pmatrix} {\vx_1}\\ \zero_{L-1} \\ {\vx_2}\end{pmatrix}^-}
  = \begin{pmatrix} \vx_1*\cc{\vx_2^-}\\ \vx_1*\cc{\vx_1^-}+\vx_2*\cc{\vx_2^-}\\ \vx_2*\cc{\vx_1^-}\end{pmatrix},
\end{align}
where we assumed for simplicity $L=L_1=L_2$.  Exactly this separation of the autocorrelation sum in \eqref{eq:3Nb} is an
sufficient structure for semi-definite relaxations to solve the phase retrieval problem or equivalently the blind
deconvolution problem. 
Note, since the cross-correlation $\va_{1,2}$ is the conjugate-time-reversal of $\va_{2,1}$, we only need $3N-3$
correlation measurements to determine $\vb$.
In \cite{WJPH17},\cite{WJPH16a} we showed the following reconstruction algorithm: 
\begin{theorem}\label{thm:4correlation}
  Let $\vx_1\in\C_{0,0}^{L_1}$ and $\vx_2\in\C_{0,0}^{L_2}$ such that the $z-$transforms  $\uX_1(z)$ and $\uX_2(z)$ do not have any
  common factors. Then $\vx^T=(\vx_1^T,\vx_2^T)\in\C^N$ with $N=L_1+L_2$ can be recovered
  uniquely up to global phase from the measurement $\vb\in\C^{4N-4}$ defined  in \eqref{eq:3Nb} by
  solving the convex program
  \begin{align}
    \find \vX\in\C^{N\times N}\quad\text{s.t.}\quad \begin{split}\Alin(\vX)=\vb\\ \vX\mgeq 0\end{split} \label{eq:fcp3N}
  \end{align}
  which has $\vX^{\#}=\vx\vx^*$ as the unique solution. 
\end{theorem}
This result can be easily reformulated as a blind-deconvolution program by knowledge of their auto-correlations.
Therefore, we only have to identify with $\vx_2 =\cc{\vh^-}$ the conjugate-time-reversal of the \emph{FIR channel} and
with $\vx_1=\vx$ the \emph{data signal}. Then the measurements are 
\begin{align}
  \vb:=  \Alin(\vX^{\#})=\begin{pmatrix} 
   \vx*\cc{\Rvx} \\ 
   \vh*\cc{\vh^-}\\
   \vx*\vh \\
 \cc{\vh^-}*\cc{\Rvx}
 \end{pmatrix}
 =\begin{pmatrix}\va_{x}\\\va_{y}\\\vy\\\cc{\vy^-}\end{pmatrix}
 =\begin{pmatrix}\va_{1,1}\\\va_{2,2}\\\va_{1,2}\\\cc{\va_{2,1}}\end{pmatrix}.
\end{align}
Hence, inserting $\va_x,\va_y$ and $\vy$ in the algorithm \eqref{eq:fcp3N} yields the solution $\vx^{\#}=e^{i\phi} (\vx,
\cc{\vh^-})$ as in \eqref{eq:reconstructionx}, if $\vx \in\C_{0,0}^{L_1}$ and $\vh
\in\C_{0,0}^{L_2}$ generate co-prime $z-$transforms $\uX(z)$ and $\uH^*(z)$. Since we have only finite many fixed $\uX$
inputs, but randomly chosen $\uH$, the probability that $\uH$ share a common zero with $\uX$ is zero.

\subsection{Simulation and Robustness} \label{sec:simulation}

\begin{figure}[t]
  \centering\includegraphics[width=0.98\columnwidth]{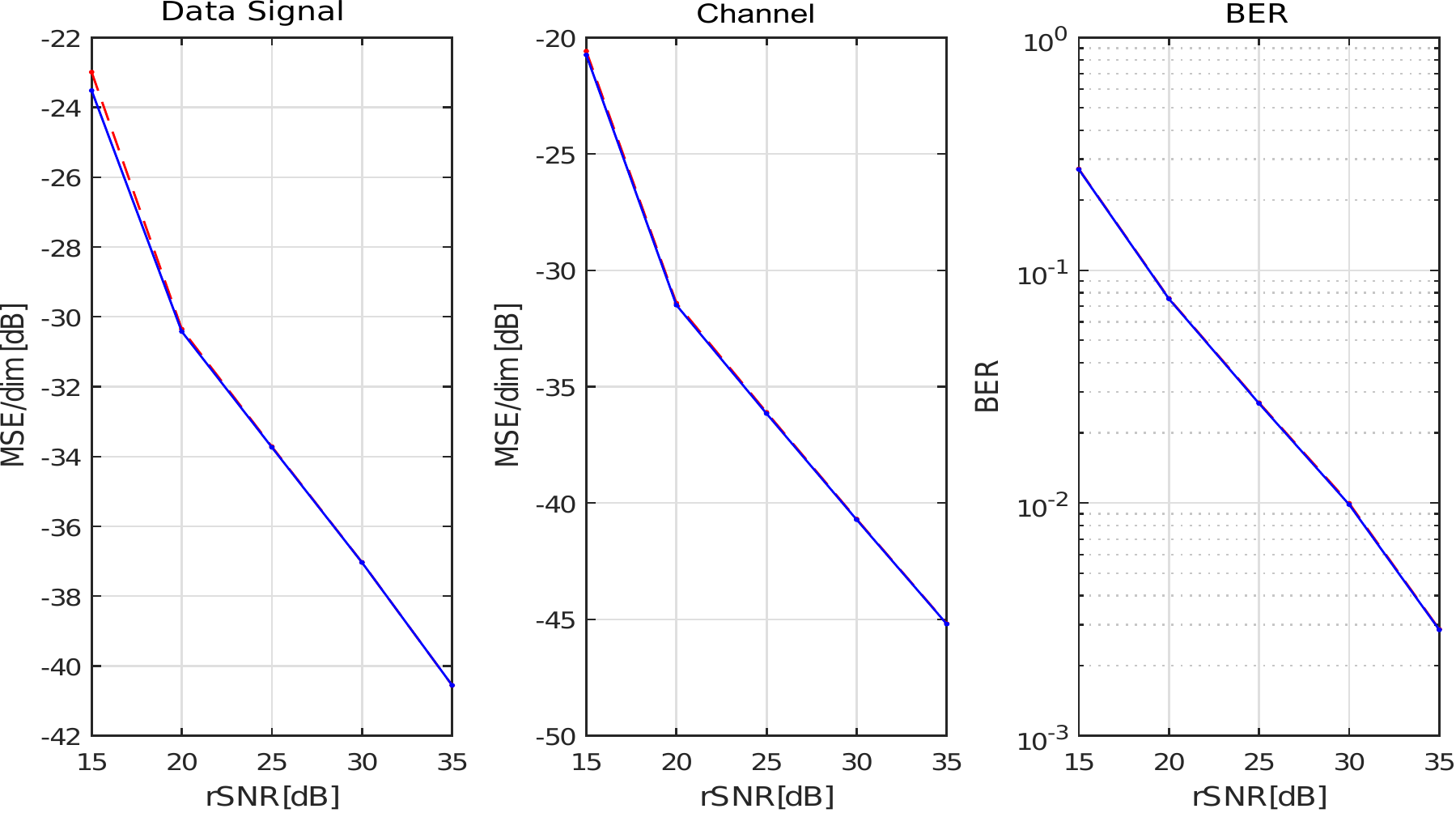}
    \caption{MSE for $13000$ runs with $L=32$ and $K=8$ of data and channel reconstruction, and BER over rSNR. 
      Red-dashed curve is with unknown and blue-solid with known energy.}\label{fig:simulation}
\end{figure}

In practice we obtain only a noisy received signal 
\begin{align}
  \vr=\vy+\vn
\end{align}
disturbed by noise vector $\vn\in\C^{L+K-1}$.  From \eqref{eq:aynoisefree} we obtain
\begin{align}
  \va_r=  \begin{pmatrix}
    -1 \\ \zero_{L-2} \\ E \\ \zero_{L-2}\\ -1\end{pmatrix} * \va_{h} + \vtn= \begin{pmatrix} -\va_h \\\zero_M \\ E
  \va_h  \\\zero_M\\
  -\va_{h} \end{pmatrix}+\vtn
= \begin{pmatrix}\va_{r,1} \\\vtn_2\\ \va_{r,3}\\ \vtn_4 \\  \va_{r,5}\end{pmatrix},\label{eq:aynoisy}
\end{align}
where $\vtn$ is the correlated noise with $\vy$ and $\vn$.
To obtain a better result for the estimation of $\va_h$ we use
\begin{align}
  \va_{\tilde{h}}=(\va_{r,1}+\va_{r,5})/2=(\va_{r,1}+\cc{\va_{r,1}^-})/2,\label{eq:average}
\end{align}
which obtains a conjugate-symmetric vector, since $\va_r$ is conjugate-symmetric. The only estimation parameter of
$\va_x$ is its energy, such that with \eqref{eq:aynoisefree} $\va_{\tx}$ is the autocorrelation of a Huffman sequence
$\vtx$ with Energy $\tilde{E}\geq 2$. Note, $E$ is actually the main-sidelobe ratio, which remains constant by scaling
of $\vx$.
Hence, we will search for the least-square solutions in \eqref{eq:fcp3N} 
\begin{align}
  \vX^{\#}:=\argmin_{\vX\mgeq 0} \Norm{\vb-\Alin(\vX)}_2^2.
\end{align}
By applying an SVD on $\vX^{\#}$ for the best rank$-1$ solution, obtains after phase reversion in
\eqref{eq:reconstructionx} our estimated signal $\vx^{\#}$.  In \figref{fig:simulation} we plotted the MSE per dimension
over the received signal-to-noise-ratio (rSNR), which scales nearly linear in dB for the signal (data) and the channel.
We also added simulation results in blue-solid if receiver knows the Energy $E$ exactly, which only effects the
reconstruction for low SNR.  In the third plot we see the uncoded Bit-error-rate (BER) over rSNR, which yields a BER of
$10^{-1}$ at 18.5dB and of $10^{-2}$ at $29$dB. To obtain better BER one might restrict the Huffman codeword  to a
smaller code, by excluding the vectors which are more likely affected by the noise in time-domain. 
%

\section{Conclusion} 
Recovering short-messages, estimating the channel and the corresponding transmit power solely from the channel output is
a challenging blind signal processing task. It is potentially relevant for next wireless technologies in the area of the
internet of things and sensor communication. We proposed here a novel scheme, based on Huffman sequences, which indeed
can provide all these tasks simultaneously in one step. It requires to solve a semidefinite program which in turn
returns the channel vector, the power at which the device transmits and after a decoding step the raw information bits.

{\it Acknowledgments.} We would like to thank Kishore Jaganathan, Fariborz Salehi, Anatoly Khina  and
Götz Pfander for helpful discussions. 

\section*{References}

\printbibliography

\end{document}